# Direct Visualization of Memory Effects in Artificial Spin Ice

Ian Gilbert[1], Gia-Wei Chern[2], Bryce Fore[1], Yuyang Lao[1], Sheng Zhang[3,4], Cristiano Nisoli[2], and Peter Schiffer[1]*

1. Department of Physics and Frederick Seitz Materials Research Laboratory, University of Illinois at Urbana-Champaign, Urbana, IL 61801, USA
2. Theoretical Division and Center for Nonlinear Studies, MS B258, Los Alamos National Laboratory, Los Alamos, NM 87545, USA
3. Department of Physics and Materials Research Institute, Pennsylvania State University, University Park, PA 16802, USA
4. Materials Science Division, Argonne National Laboratory, 9700 S. Cass Avenue, Argonne, IL 60439, USA

We experimentally demonstrate that arrays of interacting nanoscale ferromagnetic islands, known as artificial spin ice, develop reproducible microstates upon cycling an applied magnetic field. The onset of this memory effect is determined by the strength of the applied field relative to the array coercivity. Specifically, when the applied field strength is almost exactly equal to the array coercivity, several training cycles are required before the array achieves a nearly completely repeatable microstate, whereas when the applied field strength is stronger or weaker than the array coercivity, a repeatable microstate is achieved after the first minor loop. We show through experiment and simulation that this memory exhibited by artificial spin ice is due to a ratchet effect on interacting, magnetically-charged defects in the island moment configuration and to the complexity of the network of strings of reversed moments that forms during magnetization reversal.

# I. INTRODUCTION

Memory effects in hysteretic condensed matter systems have long attracted attention. These effects are seen in diverse systems, including magnetic multilayers [1-4], sheared particle suspensions [5-8], martensitic transformations [9,10], and superfluid helium in capillaries [11]. When a hysteretic system reproduces ("remembers") exactly the same microstate after being driven through a single hysteresis loop, the system is said to exhibit a return point memory [12-14]. Other systems require a number of training cycles in order to self-organize into a configuration that is replicated after subsequent cycles [6]. While memory effects are of both fundamental and applied interest, due to connections with avalanches [15,16] and the Barkhausen effect [17-19], experimental investigation of such phenomena in magnetic systems has been largely limited to collective phenomena with macroscopic consequences. Magnetic microscopy studies to determine the reproducibility of domain structures in thin film systems have yielded inconsistent results [4,20-22], which is likely a consequence of differences in material properties, thermal fluctuations, and pinning defects in the various material systems studied. Furthermore, domain structure imaging is at best a coarse-grained probe of memory at the microscopic level, since the resolution of individual spins remains difficult [23]. Some of the most detailed experimental investigations of memory effects in magnetic systems have relied on x-ray magnetic speckle measurements as a proxy for real space imaging [1-3], but again experimental resolution factors limit the theory based on these measurements to continuum models [3].

Artificial spin ices, originally designed to model frustrated magnetic systems such as spin ice [24-26], comprise arrays of interacting single-domain ferromagnetic nanoislands [27-29]. Since the nanoislands' magnetic moments have well-defined degrees of freedom and can be directly imaged by modern techniques such as magnetic force microscopy (MFM), artificial spin ice has been used to extensively study monopole-like excitations in frustrated spin systems [30-37] as well as associated cascades occurring during magnetization reversal [37-40]. Many experimental realizations of artificial spin ice possess both of the key ingredients for return point memory [14]: quenched disorder [41-43] and freedom from thermal fluctuations [27]. A recent numerical study simulated trapped colloids whose collective behavior can access an ice-like manifold [44], suggesting that artificial colloidal ice should exhibit reproducible microstates upon cycling an external field. While this work did not distinguish between return point memory (which, strictly defined, only refers to memory after the first minor loop) and more general limit cycles which may require some initial training, and while it is quite different from artificial spin ice as usually studied in magnetic systems, it strongly suggested that interesting memory effects should be seen and could be locally resolvable in artificial spin ice systems.

Here we experimentally demonstrate such memory effects, showing reproducible microstates after field cycling in artificial spin ice. For applied fields significantly stronger or weaker than the array coercivity, the final microstate is determined during the first loop, so the system exhibits nearly ideal return point memory. For applied fields approximately equal in strength to the array coercivity, interactions between effective charged defects in the moment configuration prevent the development of return point memory, and several "training" loops are required for a reproducible microstate to appear. These results are well reproduced by relaxational dynamics simulations that incorporate long-range dipolar interactions and quenched disorder. Direct MFM imaging of the island moment configuration yields unprecedented access to the microscopic details of these memory effects and allows us to explain our data in terms of a model involving the dynamics of the effective magnetic charges in artificial spin ice systems [33-35].

## II. EXPERIMENTAL DETAILS

We fabricated small arrays of square lattice artificial spin ice that completely fit within a single MFM image, which allowed us to unambiguously identify individual islands across a series of images of the same array. The arrays had closed edges (edge geometry may affect an array's response to an applied field for sufficiently-low disorder [43]), and the island spacing (360 nm), island size (220×80×25 nm), and material ($Ni_{81}Fe_{19}$) were identical to samples studied in previous works [27,45]. MFM images confirm that the islands are single-domain, with the magnetization along the islands' long axes. Although micromagnetic studies have indicated some magnetization curling may occur at the ends of the islands [46], to first order the islands may be treated as giant Ising spins, as evidenced by the agreement between our experimental data and single-spin simulations described below. We applied magnetic fields at 45 ± 0.5° to the islands (Fig. 1a), first polarizing the sample by applying a field much larger than the array's coercive field ($H_c \approx 650\,\text{Oe}$) to provide a consistent initial state, and then driving the arrays through six minor hysteresis loops by applying a magnetic field $H_a$ alternately antiparallel and parallel to the original polarizing field. After each magnetic field application, we collected MFM images of the arrays at zero field (e.g. Fig. 1b) and extracted the microscopic configuration of island moments. Since the configuration of island moments in these arrays is frozen at room temperature [27], the configuration does not change during the monotonic reduction of the magnetic field from ±$H_a$ to zero. We investigated several different sets of minor loops, each set having a different magnitude of $H_a$.

We also conducted relaxational dynamics simulations [40] of our arrays. Because detailed micromagnetic simulations of the entire arrays are too computationally expensive, we modeled our experimental data by simulating the system with individual islands approximated as point dipoles, interacting with each other via the dipolar interaction. The magnetization of the islands was described by an Ising variable $\mathbf{m}_i = s_i M_s \hat{\mathbf{e}}_i$, where $M_s$ is the saturation magnetization and $\hat{\mathbf{e}}_i$ is a unit vector specifying the orientation of the island. We employed zero-temperature relaxation dynamics for the Ising spins: an Ising spin $s_i$ was inverted if the total field acting on it, composed of external field and local field (including the fields of neighbors up to ten lattice constants away), exceeded the spin's coercivity $H_{i,c}$. Quenched disorder was incorporated in our simulations by modulating the local coercive fields according to a Gaussian distribution with a standard deviation of 5% of the mean coercive field, a disorder level which is consistent with the width of the experimental reversal curve and also provided a reasonable fit to the experimental data (Fig. 2). At each minor loop, we first inverted those spins for which the total field projected along its axis exceeded its coercive field. After updating the dipolar fields, we looked for further flippable spins and inverted those spins too. The process was repeated until there were no more flippable spins.

### III. RESULTS

The configuration of island moments in artificial square spin ice is often analyzed in terms of vertices at which four islands meet [27]. The possible configurations of these vertices are shown in the inset of Fig. 2a. As in spin ice materials [24-26,30-32], Type I and II vertices in the square lattice, which obey the ice rule (i.e., two island moments point in and two out), are lower in energy than Type III vertices, which break the ice rule (three-in/one-out or vice versa) [27] and possess an effective magnetic charge that results in behavior analogous to that of magnetic monopoles [30-35].

In Figs. 2a and b, we show the evolution of vertex populations with field cycles. The key feature of both the experimental data and simulations of the vertex populations is that they saturate after at most about four minor loops, suggesting that a repeatable microstate is established at this point. The vertex populations are an average property, however, and do not significantly add to the insight offered by previous bulk probes of memory effects. To quantify the microscopic memory in these artificial spin ice arrays, we extracted the spin overlap parameter [44],

$$q = \frac{1}{N} \sum_i s_i^{(n)} s_i^{(n-1)},$$

where $N$ is the number of islands, $s_i^{(n)} = \pm 1$ is the magnetic moment of island $i$ after minor loop $n$, and $s_i^{(n-1)}$ is the moment of the same island exactly one minor loop earlier [44]. If, after a complete minor loop, every island in an array returned to its configuration at the same point in the previous loop, $q=1$; if there is no correlation between the island configurations of loop $n-1$ and $n$, then $q \approx 0$. We can then correlate microscopic details of the development of memory effects, e.g., individual island reversals, with the overall behavior of the system, characterized by the spin overlap parameter. The experimental spin overlap parameter for one lattice is shown in Fig. 2c for different values of $H_a$. The spin overlap parameter saturates near 1.0 after a few loops. We note that the number of minor loops required to produce this memory in artificial spin ice is similar to the numerical results of Ref. 44. The spin overlap parameter increased slightly beyond six minor loops but did not reach 1.0. This can be attributed to a slight variation in the direction of the applied field in our apparatus: the experimental setup permitted alignment only to ±0.5°, and intentionally varying the field direction further decreased the saturation value of the spin overlap parameter.

The spin overlap parameter approaches saturation most slowly for $H_a \approx H_c$, in agreement with the simulations (Fig. 2d). This is confirmed in Figs. 2e and f, which show the spin overlap parameter after six minor loops as a function of applied field. In contrast, the spin overlap parameter saturates rapidly for $H_a < H_c$ and $H_a > H_c$, indicating the development of near-perfect return point memory. The development of a memory state may be understood in terms of the nucleation, dissociation, and annihilation of magnetically-charged Type III vertices during the series of field applications, as described below.

## IV. DISCUSSION

We now describe in detail how the magnetically-charged Type III vertices affect the development of memory in artificial spin ice. The interactions between island moments play a crucial role here. The long-range nature of the dipolar interaction suggests that a quantitative description must include the effects of many neighbors, as is done in our relaxational dynamics simulations; however, a qualitative understanding may be obtained by considering adjacent islands only, as shown in Fig. 3. For an island moment to reverse, the projection of the total field (applied field $\mathbf{H}_a$ plus local field $\mathbf{H}_{local}$) antiparallel to the island moment $\mathbf{m}_i$ must exceed the island's coercivity, $H_{c,i}$:

$$-\left(\mathbf{H}_a + \mathbf{H}_{local}\right) \cdot \hat{\mathbf{m}}_i > H_{c,i}.$$

If the array is initially polarized with a field in the direction that we define as positive (i.e., in the $+\hat{\mathbf{x}}+\hat{\mathbf{y}}$ direction in Fig. 4), then for the first application of $-H_a$ to the polarized array, a negative local field will assist the applied field in reversing the island moment, while a positive local field will partially cancel the applied field and inhibit island moment reversal.

Let $H_{NN}$ be the projection along an island long axis of the field from a nearest neighbor (green in Fig. 3), and similarly let $H_{NNN}$ be the projection along an island moment of the field from a next-nearest neighbor (yellow in Fig. 3). Note that $|H_{NN}| > |H_{NNN}|$. As with $H_a$, we define $H_{NN}$ and $H_{NNN}$ to be positive when they are parallel to the original direction of the island moment, i.e., when they have positive projections on the original, saturating field. The first application of $-H_a$ to the polarized array nucleates pairs of oppositely-charged Type III vertices wherever a single island of particularly low coercivity is reversed. In order for this to happen, the applied field must overcome not only the island's intrinsic coercivity, but also the local field $+2H_{NNN}$, which inhibits formation of these effective magnetic charges. The Type III vertices are then driven apart by the field through successive moment reversals of neighboring islands, leaving a string of flipped island moments joining them [34,35]. The string can grow via two different possible moves, illustrated in Fig. 4, which we name "parallel" and "perpendicular" hops.

We first consider the case of an island moment to be flipped to create a perpendicular hop (Fig. 4a,b). Adding the fields from an island's four nearest-neighbors and two next-nearest-neighbors, we find that the local field on that island is given by

$$H_{local} = 2H_{NNN} - 3H_{NN} + H_{NN} = -2(H_{NN} - H_{NNN}) < 0.$$

Note that, for this case, the local field on the island to be flipped is negative, so the local field adds to the applied field, helping to induce the perpendicular hop.

We now consider the case of an island moment to be flipped to create a parallel hop (Fig. 4c,d). The local field on an island moment to be flipped to create a parallel hop is

$$H_{local} = H_{NNN} - H_{NNN} + 2H_{NN} - 2H_{NN} = 0.$$

For a parallel hop, the local field is zero, so it has no effect on reversing the island moment. The string can thus grow through either parallel or perpendicular hops until it terminates when the Type III vertices at the ends encounter islands of particularly high intrinsic coercivity and become pinned.

When the applied field is reversed ($+H_a$), the magnetically-charged Type III vertices will be driven back toward each other, erasing the string as they go. Since the local field for a parallel hop is zero, the parallel hops can be retraced without difficulty. The local field for perpendicular hops, parallel to the applied field in the previous step, now partially cancels the applied field and may prevent the Type III vertices from retracing the perpendicular hops, depending on the coercivity of the individual island. The local field thus provides a ratchet mechanism that supports the growth of the strings when $-H_a$ is applied, but impedes their erasure when $+H_a$ is applied.

In Fig. 5, we demonstrate the dynamics of the magnetic charges for $H_a$ well below $H_c$, showing an example of our experimental data taken from MFM images and displayed to highlight the flipped moments and Type III vertices. In this case, the first application of $-H_a$ after polarization (Fig. 5a) nucleates only isolated pairs of Type III vertices, so their associated strings rarely cross. When the field is switched back and forth, the charged Type III vertices simply move back and forth at the ends of their respective strings but do not significantly interact with each other. They return to the same pinning site after each field application, and so the moment configuration will be reproduced after the first minor loop, as can be seen in Fig. 5. Similar behavior is observed in the simulation results, presented in Fig. 6. Consequently, the spin overlap parameter saturates rapidly, as is seen in both experiment $H_a/H_c = 0.949$ in Fig. 2a) and simulation (Fig. 2b), and the sample exhibits near-perfect return point memory.

In Fig. 7, we demonstrate the dynamics of the magnetic charges for $H_a \approx H_c$, again showing experimental data. For $H_a \approx H_c$, the Type III vertices and strings are no longer isolated. During the first application of $-H_a$, the Type III vertices annihilate with oppositely-charged Type III vertices from other strings, and strings can merge or cross. The result is a *network* of inverted island moments, shown in Fig 7a, rather than just sparse tracks. This leads to an entirely different regime of magnetic charge annihilation during the subsequent application of $+H_a$ (Fig. 7b). Since the Type III vertices populate the boundaries of a complex network of inverted moments, when the field is reversed again, the Type III vertices can again annihilate with oppositely-charged Type III vertices other than the ones from which they originally dissociated. In particular, a charge that had tracked a step-like string no longer has to annihilate with the oppositely-charged vertex from which it originally dissociated, but can now travel back along an easier parallel path to annihilate with a different one. The dynamics is collective and complex, with the trajectories of the Type III vertices dictated by network's topology, interactions with other Type III vertices, and the spatial distribution of local disorder. The network of strings allows for more freedom during each field reversal as the complicated magnetic charge and string interactions react to the *new initial conditions* produced after each minor loop, delaying the development of a repeatable memory state. In this case, the behavior of the system cannot be properly termed return point memory, but may be more appropriately classified as a limit cycle. This explains the delayed saturation of the spin overlap parameter for $H_a/H_c = 1.005$ in Fig. 2a and b, and the larger differences among the panels of Fig. 7 compared to those of Fig. 5. Nevertheless, the differences between the moment configurations decrease as the number of minor loops increases. The Type III vertices eventually find the most favorable configurations, and so the spin overlap parameter saturates (within experimental uncertainty) after about four minor loops, similar to Ref. 44. Again, our simulations produce similar results, shown in Fig. 8. The dynamics of the magnetic charges is approximately symmetric in applied field, since at still higher applied magnetic fields, $H_a$ well above $H_c$, nearly all the island moments are reversed by the first application of $-H_a$, leaving only a sparse distribution of strings (now comprised of *unflipped* islands) behind.

## V. CONCLUSION

Our results demonstrate how artificial spin ice can be used to examine the microscopic basis of other phenomena associated with hysteretic systems. We envisage several possibilities for future works. For example, investigating thermally active artificial spin ice [47] would further permit the microscopic investigation of non-repeatable [20-22] as well as repeatable [4,18] Barkhausen effects. Here we have demonstrated how applied field can tune regimes, from unidirectional, ratchet-like behavior to more complex dynamics on a network. Nontrivial geometries of artificial spin ice [48,49] can expand this control, potentially allowing such exotic phenomena as chiral magnetic ratchets, as well as magnetricity-based devices [50]. Furthermore, the coupling of these sources and sinks of the H field to 2D systems of correlated electrons could lead to control of quantum states, such as superconducting vortices and antivortices [51-53], through the manipulation of the magnetic charges, suggesting numerous paths for future studies.

## ACKNOLWEDGEMENTS

This work was primarily funded by the US Department of Energy, Office of Basic Energy Sciences, Materials Sciences and Engineering Division under grant no. DE-SC0010778. Electron beam lithography was supported by the National Nanotechnology Infrastructure Network. The work of G.-W. Chern and C. Nisoli was carried out under the auspices of the US Department of Energy at LANL under contract no. DE-AC52-06NA253962. We thank Liam O'Brien for assistance with sample fabrication and Karin Dahmen and James Sethna for useful discussions.

_________________________________________

*pschiffe@illinois.edu

# Figures

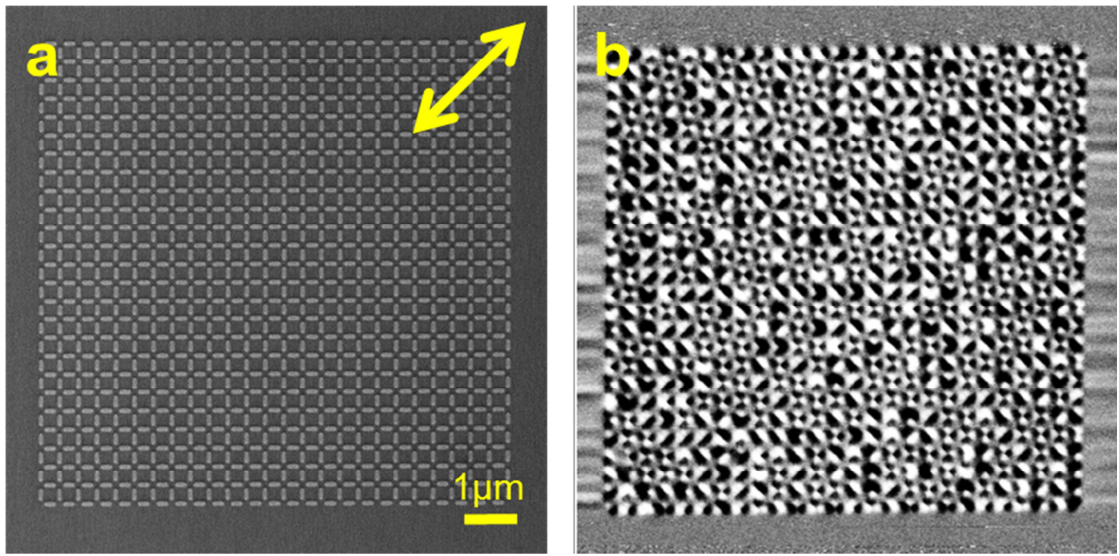

FIG. 1 (color). Artificial spin ice for memory measurements. Panel (a) shows a scanning electron micrograph of the 360 nm square lattice, and panel (b) shows a magnetic force microscope image of the same lattice. The black/white contrast in (b) indicates the north and south magnetic poles of the nanomagnets. The double-headed arrow in (a) denotes the axis along which magnetic fields were applied.

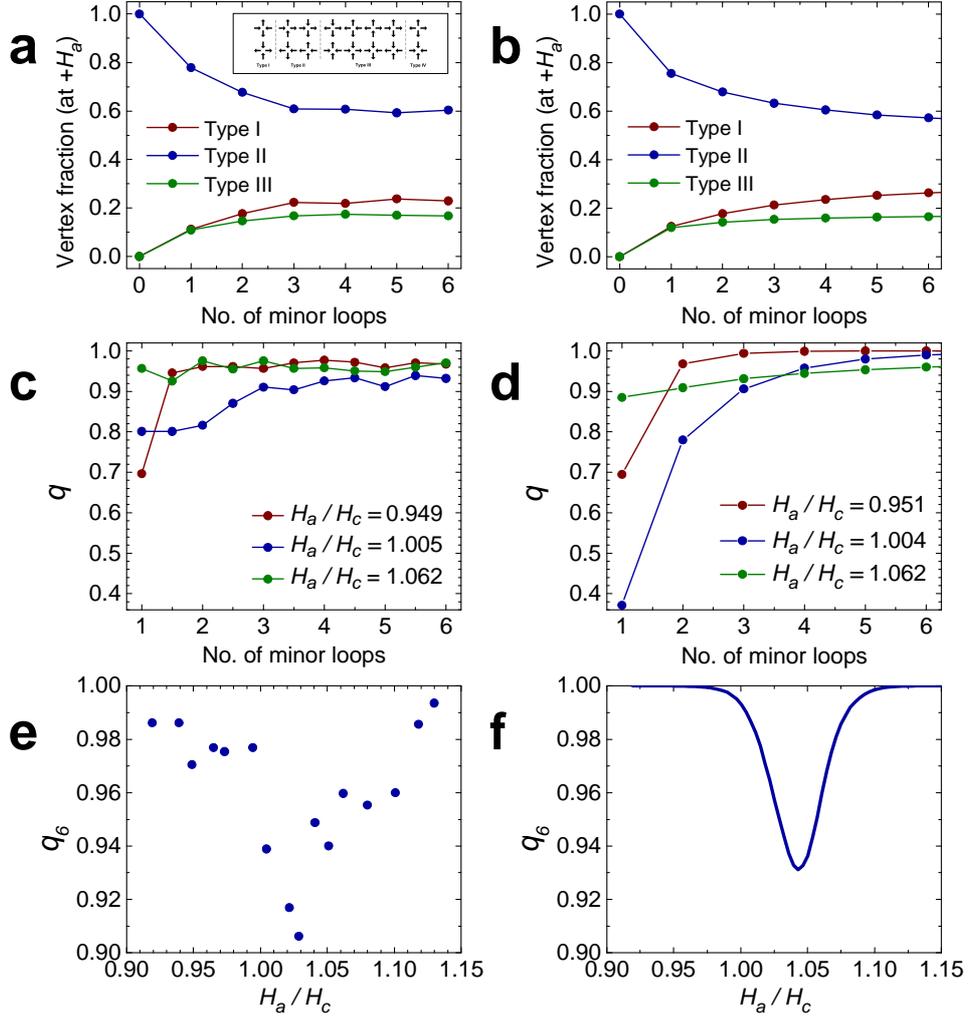

FIG. 2 (color). The development of memory in artificial square spin ice. The populations of the different types of vertices (enumerated in the inset of (a)) are shown as a function of number of minor loops for experiment (a) and simulation (b). The evolution of the spin overlap parameter $q$ with repeated cycles of field is shown in panels (c) and (d) for experiment and simulation, respectively. The spin overlap parameter saturates more slowly in the intermediate field regime ($H_a / H_c = 1.005$) than for $H_a < H_c$ and $H_a > H_c$. Panels (e) and (f) show the spin overlap parameter after six minor loops ($q_6$) as a function of $H_a / H_c$. Experiment (e) and simulation (f) both show a minimum just above $H_a / H_c = 1$.

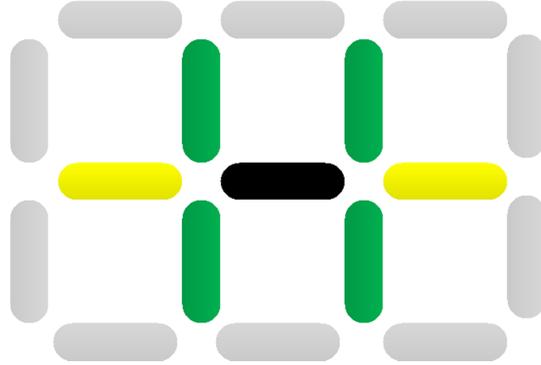

FIG. 3 (color). Near neighbors in artificial spin ice. When calculating the local field produced on an island by its neighbors, we here consider only the adjacent islands, i.e. nearest- and next-nearest-neighbors. For example, for the black island above, we would only consider the field $H_{NN}$ from each of the four nearest neighbors (green) and the field $H_{NNN}$ from each of the two next-nearest-neighbors (yellow). Note that micromagnetic simulations show the interactions between the central black island and the NNN yellow islands are more than three times stronger than the interactions with the horizontal gray islands directly above and below the black island, even though both pairs are separated by the same distance. Consequently, we include only the yellow NNN islands in our qualitative description.

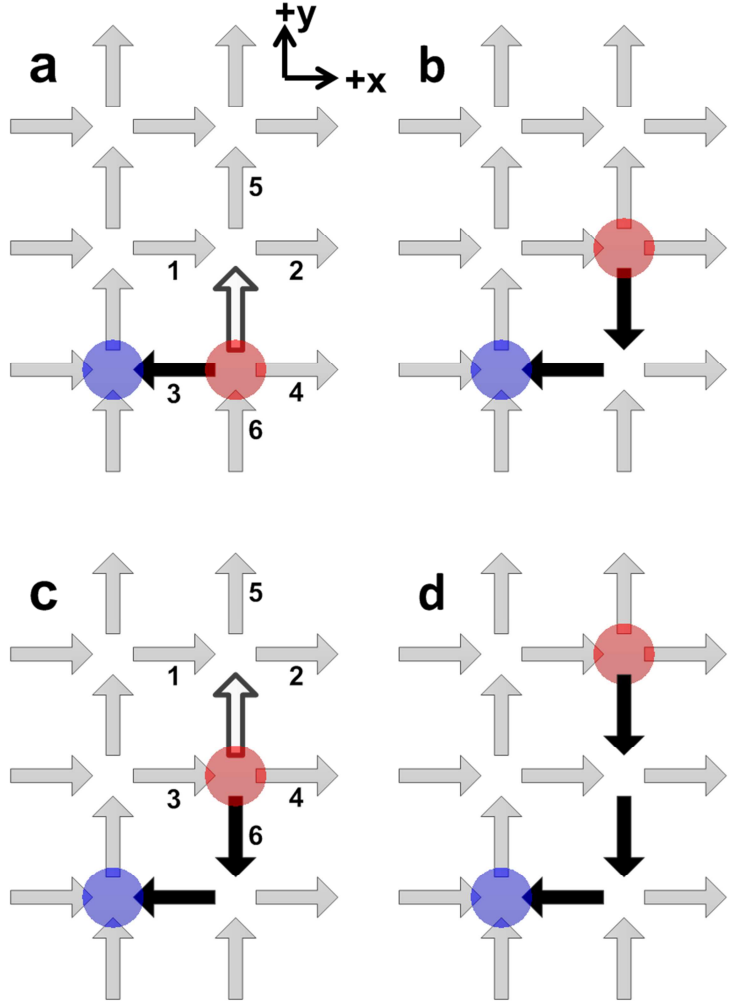

FIG. 4 (color). Hopping possibilities for a magnetically-charged Type III vertex lengthening a string of reversed island moments. The applied field in all panels is in the $-\hat{\mathbf{x}}-\hat{\mathbf{y}}$ direction. The outlined island in (a) reverses to form a perpendicular hop, shown in (b). Of the nearest neighbors (islands 1-4) of the island outlined in (a), the fields from 1, 3, and 4 all add to the applied field (since they are head-to-head or tail-to-tail with the outlined island), while the field from 2 partially cancels it (since it is already aligned head-to-tail with the outlined island moment). Similarly, both of the next-nearest-neighbors (islands 5 and 6) are already aligned head-to-tail with the outlined island, producing fields that inhibit reversal. Adding local field contributions from all six neighbors, we find that $H_{local} = \sum_{i=1}^{6} H_i = -H_{NN} + H_{NN} - H_{NN} - H_{NN} + H_{NNN} + H_{NNN} = -2(H_{NN} - H_{NNN})$. The outlined island in (c) reverses to form a parallel hop, shown in (b). Adding the local field contributions as before, we find that

$H_{local} = \sum_{i=1}^{6} H_i = -H_{NN} + H_{NN} + H_{NN} - H_{NN} + H_{NNN} - H_{NNN} = 0$. Note that while only vertical island moment reversals are shown here, vertical and horizontal reversals occur with equal frequency since the external magnetic field is applied at 45° to the islands.

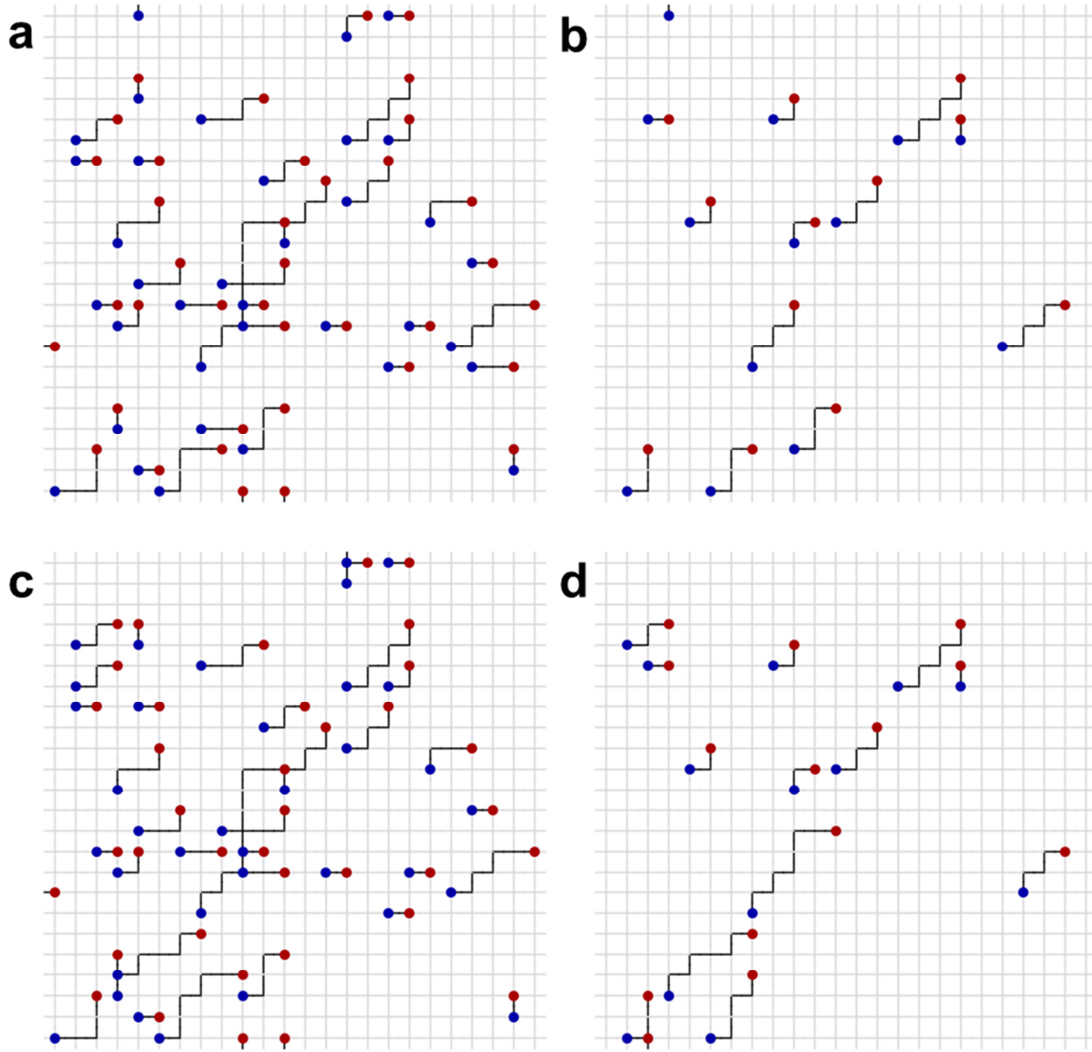

FIG. 5 (color). Snapshots of the experimental string configuration in the regime $H_a$ well below $H_c$ (taken from MFM images for $H_a / H_c = 0.919$). The configuration is shown after the first $-H_a$ (a), the first $+H_a$ (b), the second $-H_a$ (c), and the second $+H_a$ (d). Positively (negatively) charged Type III vertices are represented by red (blue) dots, gray lines represent island moments aligned by the original polarizing field or $+H_a$, and black lines represent moments reversed by $-H_a$. When the field is brought back to $+H_a$ (b), Type III vertices can only partially retrace their associated string. From this history, the configurations at further inversions are determined, so (c) and (d) are almost identical to (a) and (b), thus yielding return point memory.

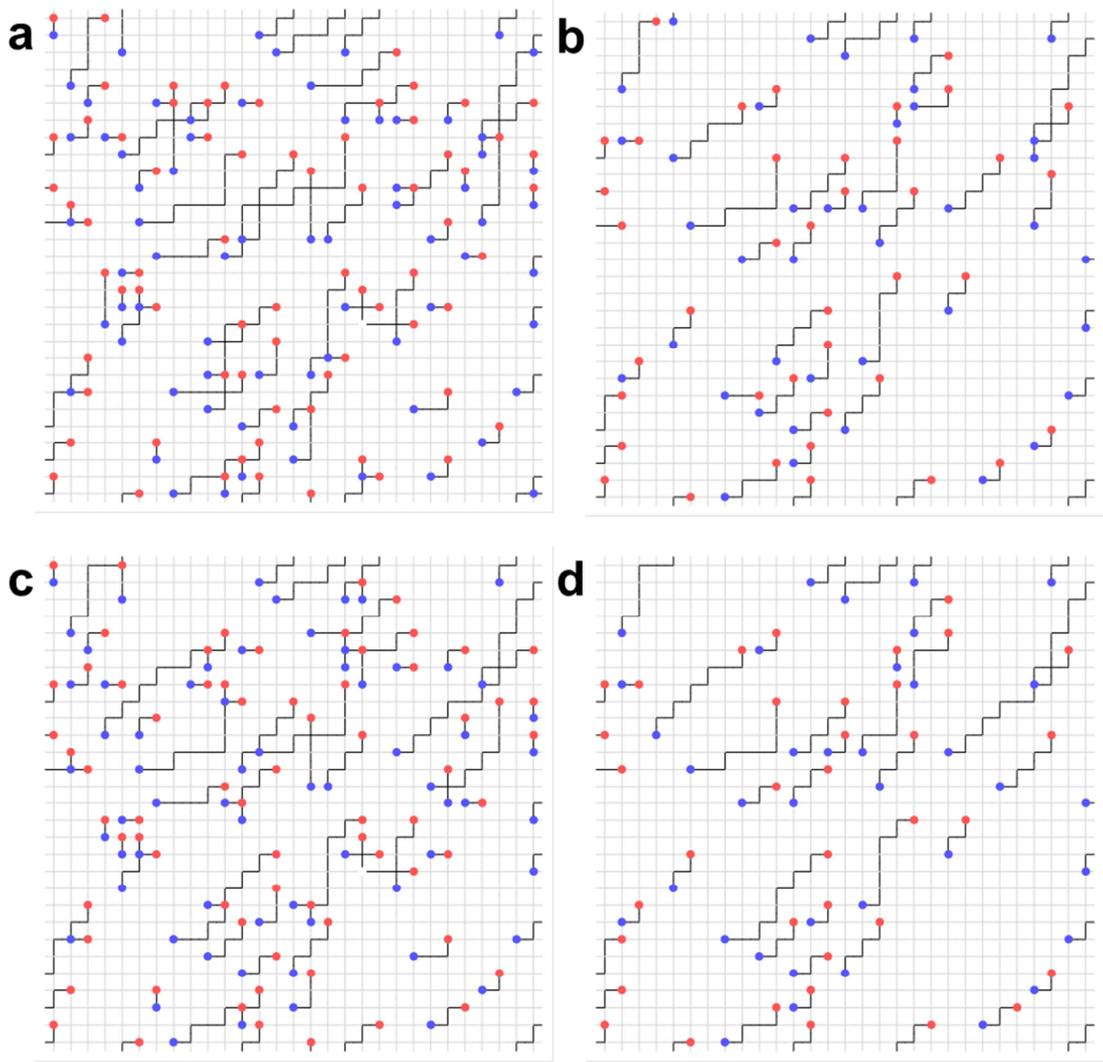

FIG. 6 (color). Snapshots of the simulated string configuration of artificial square spin ice arrays with parameters similar to those of our experimental data, with $H_a = 0.91 H_c$. The configuration is shown after the first $-H_a$ (a), the first $+H_a$ (b), the second $-H_a$ (c), and the second $+H_a$ (d). Positively (negatively) charged Type III vertices are represented by red (blue) dots, gray lines represent island moments aligned by the original polarizing field or $+H_a$, and black lines represent moments reversed by $-H_a$. As in Fig. 5, we see that there are strong similarities between island moment configurations one loop apart (i.e., compare (a) with (c) and (b) with (d)).

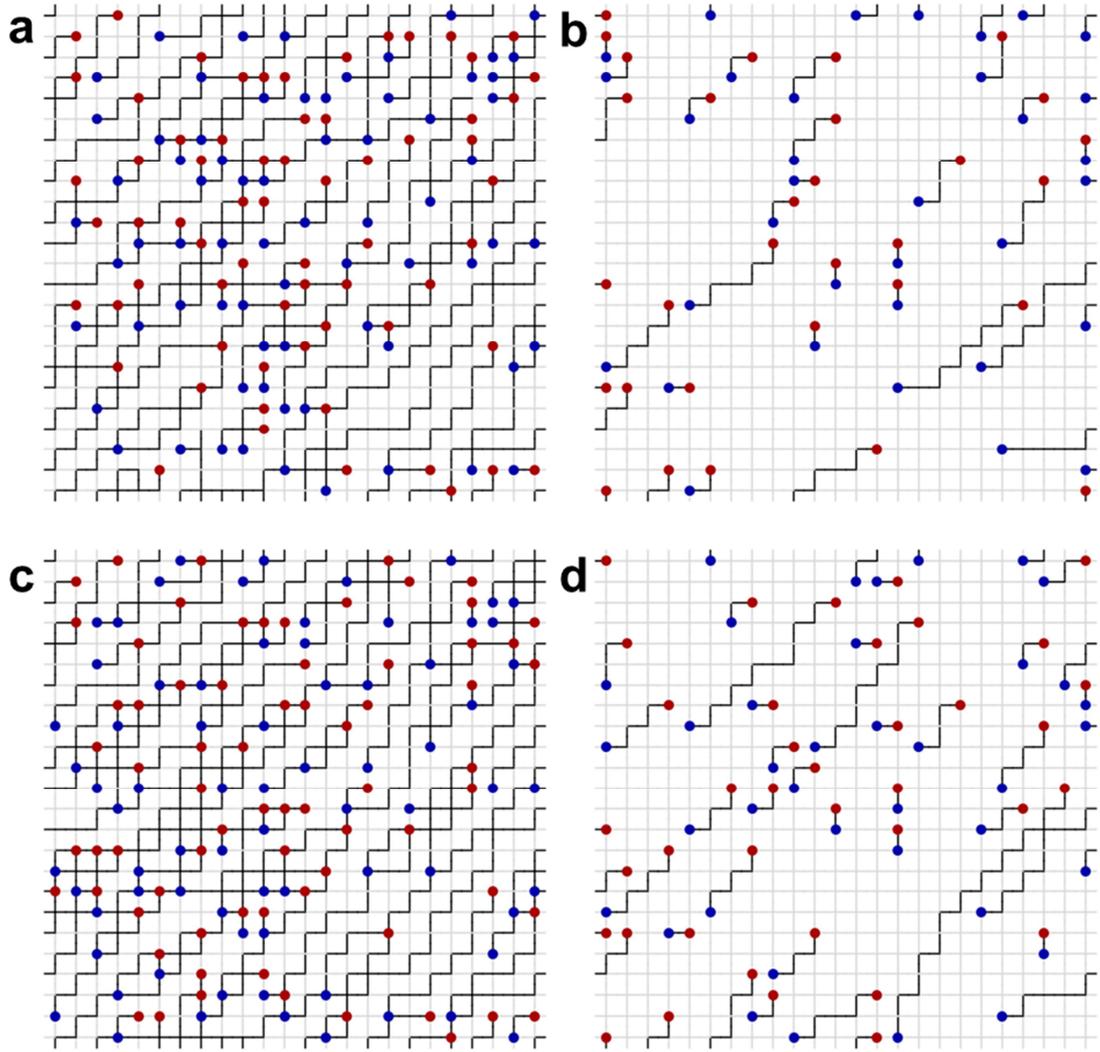

FIG. 7 (color). Snapshots of the experimental string configuration in the regime $H_a \approx H_c$. These data are taken from MFM images for $H_a / H_c = 1.005$. The configuration is shown after the first $-H_a$ (a), the first $+H_a$ (b), the second $-H_a$ (c), and the second $+H_a$ (d). Positively (negatively) charged Type III vertices are represented by red (blue) dots, gray lines represent island moments aligned by the original polarizing field or $+H_a$, and black lines represent moments reversed by $-H_a$. The Type III vertices can choose new paths to annihilate, leaving the tracks in (b) when the field is restored to $+H_a$. This asymmetry between dissociation and annihilation means that the configurations in the subsequent loop (after the second $-H_a$ (c) and the second $+H_a$ in (d)) are somewhat different, delaying the development of a repeatable limit cycle state.

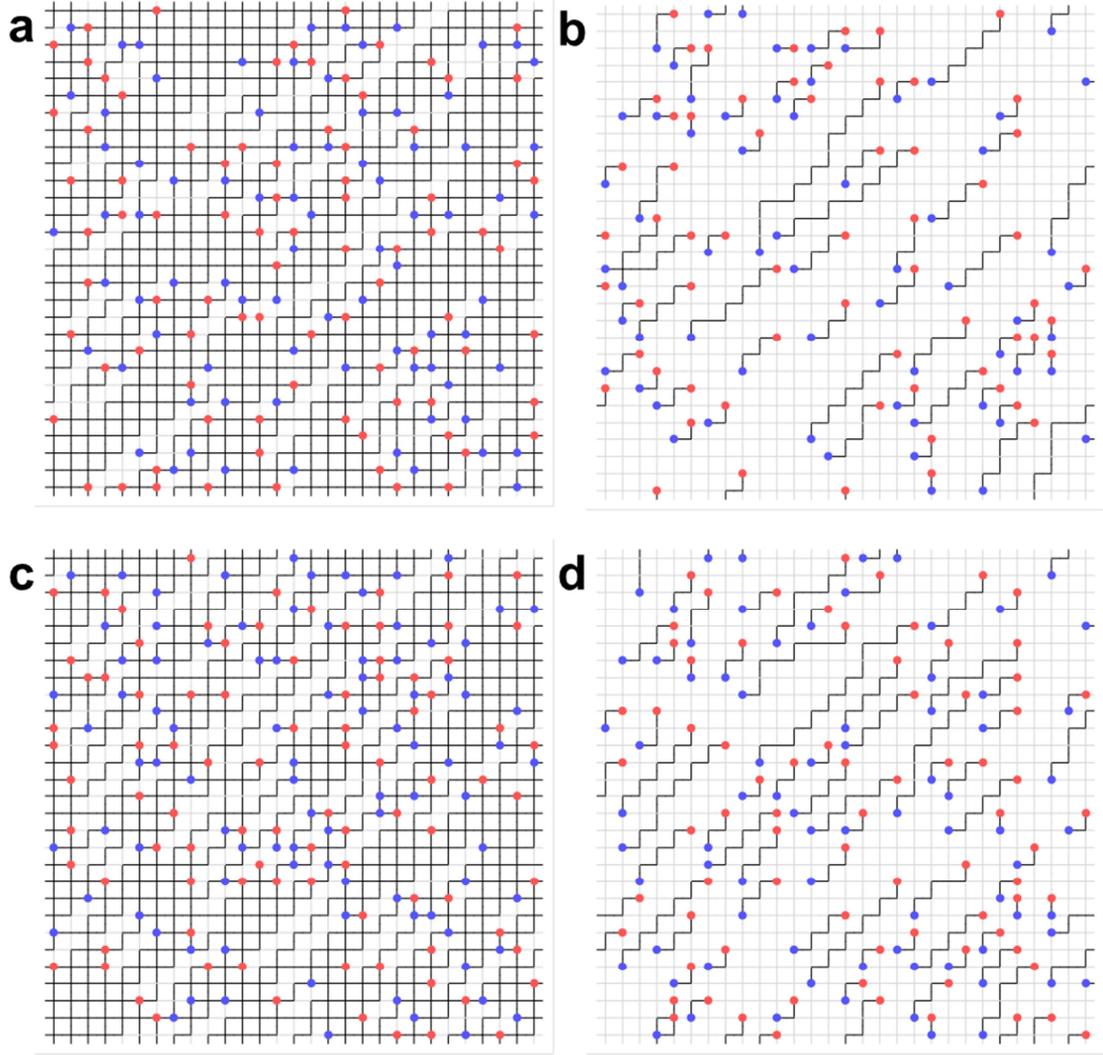

FIG. 8 (color). Snapshots of the simulated string configuration of artificial square spin ice arrays with parameters similar to those of our experimental data, with $H_a = 1.03 H_c$. The configuration is shown after the first $-H_a$ (a), the first $+H_a$ (b), the second $-H_a$ (c), and the second $+H_a$ (d). Positively (negatively) charged Type III vertices are represented by red (blue) dots, gray lines represent island moments aligned by the original polarizing field or $+H_a$, and black lines represent moments reversed by $-H_a$. As in Fig. 7, we see some similarities between the island moment configurations in (a) and (c) and in (b) and (d), but not as much as in Figs. 5 and 6. The stronger applied field produces a denser network of interacting magnetic charges, and these interactions (including annihilation of opposite charges as well as long-range Coulomb interactions [54,55]) inhibit the development of perfect memory.